\documentclass[twocolumn,aps,superscriptaddress,floatfix,showpacs]{revtex4}
\usepackage{graphics,epsfig,graphicx}  
\begin{document}

\title{The variable phase method used to calculate and correct scattering lengths}

\author{H. Ouerdane}
\affiliation{Department of Computing Science, University of Glasgow, Glasgow G12 8QQ, Scotland, UK}
\author{M. J. Jamieson}
\affiliation{Department of Computing Science, University of Glasgow, Glasgow G12 8QQ, Scotland, UK}
\author{D. Vrinceanu}
\affiliation{Harvard-Smithsonian Center for Astrophysics, 60 Garden Street, Cambridge, MA 02138, USA}
\author{M. J. Cavagnero}
\affiliation{Department of Physics and Astronomy, University of Kentucky, Lexington, KY 40506-0055, USA}

\begin{abstract}
It is shown that the scattering length can be obtained by solving a Riccati equation derived from variable phase theory. Two methods of solving it are presented. The equation is used to predict how long range interactions influence the scattering length, and upper and lower bounds on the scattering length are determined. The predictions are compared with others and it is shown how they may be obtained from secular perturbation theory.
\end{abstract}

\maketitle

\section{Introduction}

It is well established that knowledge of the scattering length of a pair of colliding atoms is important in the interpretation of observations of the behaviour of cold trapped ensembles of such atoms \cite{KET96,JUL93,WEI99}. Scattering lengths are usually calculated by solving the single-channel radial Schr\"odinger equation to find the phase shifts for several small values of the wavenumber of relative motion and extrapolating from effective range expansions \cite{MOT65}, or by solving the zero-energy Schr\"odinger equation and evaluating quadratures \cite{GUT84}.
 
The Schr\"odinger equation is a second order differential equation. We shall show another method of evaluating the scattering length in which we solve the first order equation, a Riccati equation, of the variable phase approach to potential scattering \cite{CAL67}. Solving the equation is not simple and we shall suggest and illustrate two methods that circumvent the inherent difficulties. A differential equation can be obtained for the effective range \cite{CAL67} but it is not amenable to numerical solution.

In obtaining the scattering length from the solution of the Schr\"odinger or Riccati equation, we must obtain solutions at infinite values of the interatomic separation, $R$. In practice we stop the solution at some value $R_{\rm c}$, but $R_{\rm c}$ must be chosen very large, and such choices increase computation time and accumulated error. We shall show how the Riccati equation can be used to predict corrections, to be applied to the {\it calculated} scattering length, that compensate for stopping the calculation at finite distance $R_{\rm c}$; such corrections enable us to make calculations with smaller values of $R_{\rm c}$ while maintaining desired accuracy \cite{ORL99,JAM03,CAV94}.

The solution of a low energy Schr\"odinger equation may be corrected for the influence of a long range interaction at separations exceeding $R_{\rm c}$ by perturbation theory. However the perturbation expansion is made in terms of the ratio of the potential energy to the kinetic energy and becomes invalid as the energy of the collision is reduced (as is necessary in the usual methods to obtain a scattering length). This problem may be avoided by use of secular perturbation theory \cite{CAV94}. We shall discuss our predicted corrections in the light of secular perturbation theory and also compare them with correction formulae that have been obtained elsewhere \cite{HIN71,MAR94,SZM95}.

\section{Variable Phase Theory}

The $s$-wave Schr\"odinger equation for a pair of colliding atom is:

\begin{equation} \label{eq1}
  \left[\frac{\displaystyle {\rm d}^2}{\displaystyle {\rm d}R^2} - V(R) + k^2\right] y(k;R) = 0,
\end{equation}

\noindent where $V(R)=2\mu {\mathcal V}(R)/\hbar^2$, ${\mathcal V(R)}$ being the interaction potential, $\mu$ the reduced mass, $\hbar$ the rationalised Planck's constant, $k$ the asymptotic (for large separation, $R$) wavenumber of relative motion and $R^{-1} y(k;R)$ the wavefunction. The phase shift $\delta_k$ is obtained from the wavefunction at large $R$:

\begin{equation} \label{eq2}
  y(k;R) \sim \sin\left(kR+\delta_k\right).
\end{equation}

\noindent Suppose that $\delta_k(R_{\rm c})$ is the phase shift appropriate to the truncated potential ${\mathcal V}(R)H(R_{\rm c}-R)$ where $H$ denotes Heaviside's unit step function. The phase shift $\delta_k$ is the limit of $\delta_k(R_{\rm c})$ as $R_{\rm c}\rightarrow\infty$. The phase shift $\delta_k(R_{\rm c})$ is the phase shift that is determined numerically when the potential is assumed to be negligible at $R>R_{\rm c}$; it has long been studied, is called the variable phase function and satisfies a first order differential equation in $R_{\rm c}$ \cite{CAL67}.

The potential for a pair of scattering alkali atoms is dominated at long range by the van der Waals interaction $-C_6R_{\rm c}^{-6}$ and the well known problem of finding ultra-low energy phase shifts from Eqs.~(\ref{eq1}) and (\ref{eq2}) is that $R_{\rm c}$ must be chosen very large to ensure that $-C_6R^{-6}$ has negligible magnitude compared to $\hbar^2k^2/2\mu$ when $k$ itself is very small.

The scattering length $a$ is the limit of $-k^{-1}\tan \delta_k$ for vanishing $k$ \cite{MOT65}. The scattering length $a(R_{\rm c})=a_{\rm c}$ is the corresponding limit of $-k^{-1}\tan \delta_k(R_{\rm c})$ and $a_{\rm c}\rightarrow a$ as $R_{\rm c}\rightarrow \infty$. Calogero \cite{CAL67} derived the Riccati equation in $R_{\rm c}$ satisfied by $a(R_{\rm c})$ or $a_{\rm c}$. We simplify the notation by dropping the subscript c and note that Calogero's definition of $a(R)$ is opposite in sign to that used in the effective range expansion \cite{MOT65}. We see that the Riccati equation is

\begin{equation} \label{eq3}
  \frac{\displaystyle {\rm d}a(R)}{\displaystyle {\rm d}R} = \left[R-a(R)\right]^2 V(R).
\end{equation}

\noindent Hence we can find the scattering length by solving Eq.~(\ref{eq3}) over the interval $[0,R_{\rm c}]$ with initial condition $a(0)=0$ where $R_{\rm c}$ is chosen sufficiently large.

\section{Numerical Methods}

The numerical solution of Eq.~(\ref{eq3}) is non-trivial. The function $a(R)$ contains poles that correspond to the bound states supported by the potential as we illustrate in figure 1 for the test case described in section 3.3. Finite difference methods such as the Runge-Kutta method are unsuitable. Two possible ways to solve Eq.~(\ref{eq3}) are to change its variables, and to solve a closely related equation by the log-derivative method.\\

\begin{figure}[!rh]
\centering
\scalebox{0.33}{\rotatebox{0}{\includegraphics*{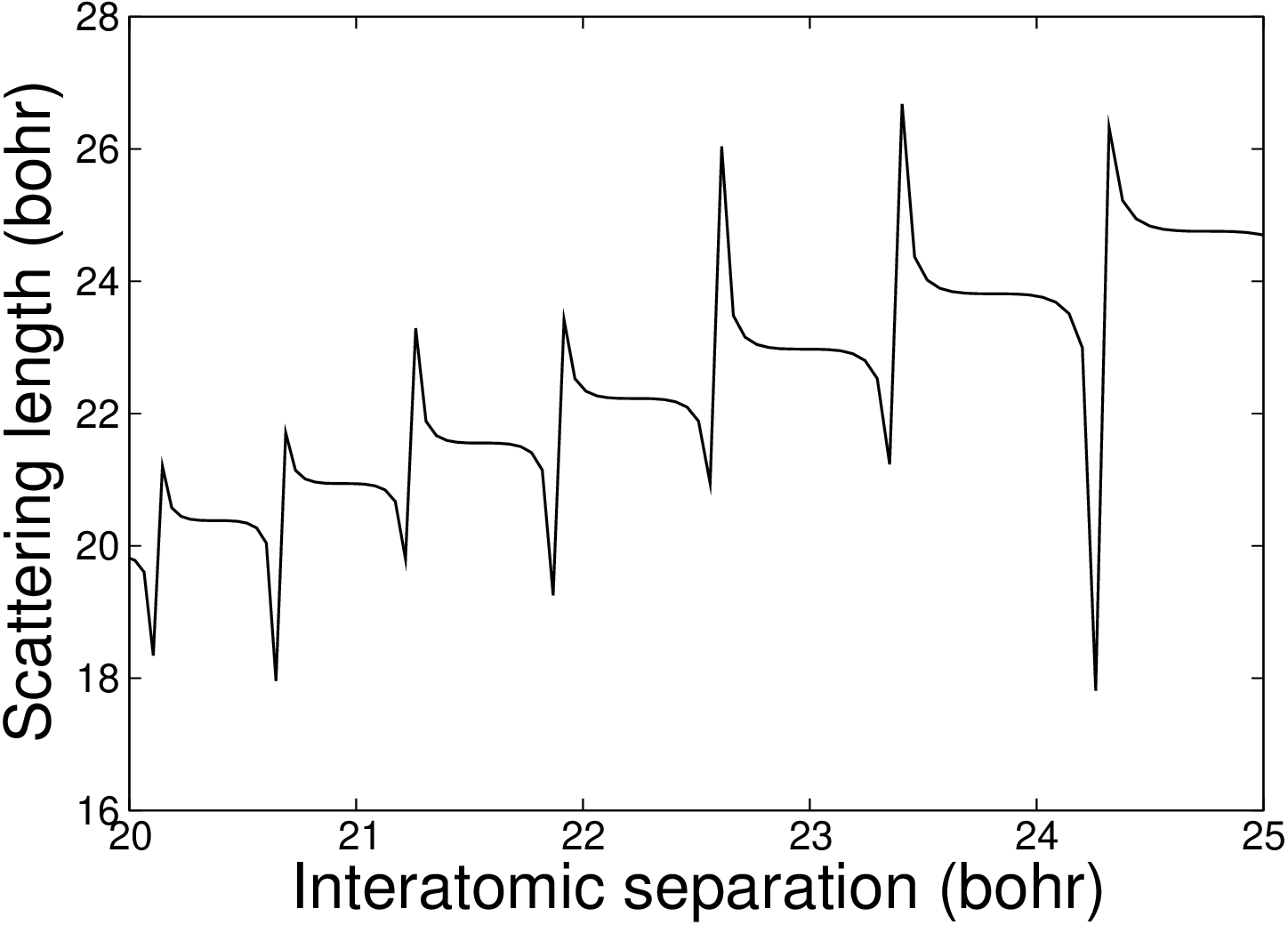}}}
\caption{The accumulated scattering length, $a(R)$, as a function of interatomic separation $R$.}
\end{figure}

\subsection{Change of Variables}

We make the change of variables\footnote{It is possible to make only the change of variable for $a(R)$ given in Eq.~(\ref{eq4}) but, with such a change, more computational time is necessary.}

\begin{equation} \label{eq4}
  a(R) = \tan \theta(R),
\end{equation}

\noindent and

\begin{equation} \label{eq5}
  R = \tan \phi(R),
\end{equation}

\noindent to obtain the equation

\begin{equation} \label{eq6}
  \frac{\displaystyle {\rm d}\theta(\phi)}{\displaystyle {\rm d}\phi} = \sec^4\phi\sin^2\left[\theta(\phi)-\phi)\right] V[\tan(\phi)],
\end{equation}

\noindent which we solve over a range $[0,\phi_{\rm c}]$ by the Runge-Kutta method \cite{CHE61}. The value of $\phi_{\rm c}$ corresponds to $R_{\rm c}$ and is close to $\pi/2$.

\subsection{Log-derivative Method}

In Eq.~(\ref{eq3}) we substitute $u(R)=\left[R-a(R)\right]^{-1}$ to obtain the equation

\begin{equation} \label{eq7}
  \frac{\displaystyle {\rm d}u(R)}{\displaystyle {\rm d}R}~+u^2(R) - V(R) = 0
\end{equation}

\noindent which is the Riccati equation for the log-derivative of the radial wavefunction at zero energy \cite{JOH73,JOH77}. The substitution is also inferred by the asymptotic form, $R-a$, of the radial wavefunction at large $R$. The log-derivative method can be evaluated by the propagator \cite{MAN93,FRI94}

\begin{widetext}
\begin{equation} \label{eq8}
\left[ 1- hu(R+h)+~\frac{\displaystyle 1}{\displaystyle 3}~h^2V(R+h)\right]^{-1} + \left[ 1+ hu(R-h)+~\frac{\displaystyle 1}{\displaystyle 3}~h^2V(R-h)\right]^{-1} = 2\left[1 -~\frac{\displaystyle 1}{\displaystyle 6}~h^2V(R)\right]^{-1} \left[1+~\frac{\displaystyle 1}{\displaystyle 2}~h^2 V(R)\right],
\end{equation}
\end{widetext}

\noindent with the initial condition that $u(0)$ is very large, where $h$ is the step length; the local truncation error is ${\mathcal O}(h^6)$ \cite{FRI94}. The success of the log-derivative method in treating the poles of $u(R)$ is attributable to its derivation from finite difference approximations to the wavefunction and its derivative \cite{MAN93,FRI94,MAN95}; it is an example of a symplectic integrator in which the distance coordinate and its canonical partner are advanced simultaneously \cite{MAN95}.
\subsection{Numerical examples}

We tested the numerical methods of sections 3.1 and 3.2 by evaluating the scattering length for a pair of caesium atoms, each of mass $2.422\times 10^5$ atomic units (a.u) interacting via the model potential described by Gribakin and Flambaum \cite{GRI93}

\begin{equation} \label{eq9}
  {\mathcal V}(R) = \alpha R^{\beta} \exp(-\gamma R) - \left( C_6R^{-6} + C_8^{-8} + C_{10}R^{-10}\right)f(R)
\end{equation}

where

\begin{equation} \label{eq10}
  f(R) = H(R - R') + H(R' - R)\exp\left[-(R'/R-1)^2\right]
\end{equation}

\noindent with $\alpha=0.0008$, $\beta=5.53$, $\gamma=1.072$, $C_6=7020$, $C_8=1.1\times 10^6$, $C_{10}=1.7\times 10^8$ and $R'=23.165$, all in atomic units. As is the case in Numerov's method the local truncation error in the log-derivative method is approximately proportional to $h^6|V(R)|^3$ \cite{BLA67}. To economise on computation and reduce possible truncation error we doubled the step length whenever the local truncation error had reduced by approximately a factor of 2 \cite{JAM03,COT95} when using the log-derivative method; the propagator in Eq.~(\ref{eq8}) is easily modified to accomodate the change in step length. When solving the first order equation, Eq.~(\ref{eq6}), we used the Runge-Kutta method \cite{CHE61} with a self-adjusting step length. Both methods yielded the same scattering length as expected which was accurate to 7 significant figures with $R_{\rm c}=40000$ bohr in agreement with the findings of Marinescu \cite{MAR94}. The variable phase function is illustrated by the curve for $a_{\rm c}$ in figure 2; the last pole is located at around 100 bohr and for $R>$100 bohr the function is smooth and approaches its limit ultimately from above.

\section{Corrections to the Scattering Length}

Eq.~(\ref{eq3}) illustrates clearly the effect of neglecting the potential for $R>R_{\rm c}$ since the equation can be recast as
\begin{equation} \label{eq11}
  a = \int_0^{R_{\rm c}}\left[R - a(R)\right]^2V(R) {\rm d}R + \int_{R_{\rm c}}^{\infty}\left[R - a(R)\right]^2V(R) {\rm d}R.
\end{equation}

\noindent The first part of this equation is the scattering length $a_{\rm c}=a(R_{\rm c})$ calculated by any numerical method in which the computation is stopped at separation $R_{\rm c}$. The second part is the contribution made by the long range interaction; the choice of $R_{\rm c}=40000$ bohr in the calculations of section 3.3 was needed to ensure that the second part was sufficiently small to give seven figure accuracy in the scattering length.

\subsection{The First Order Correction}

The correction is

\begin{equation} \label{eq12}
  E_{\rm c}= E(R_{\rm c}) = \int_{R_{\rm c}}^{\infty}\left[R - a(R)\right]^2V(R) {\rm d}R.
\end{equation}

\noindent We wish to extract the term that is of first order in the potential strength. We evaluate the quadrature, Eq.~(\ref{eq12}), by parts, replacing ${\rm d}a/{\rm d}R$ by Eq.~(\ref{eq3}) whenever it occurs, thus increasing the order each time. We find

\begin{eqnarray} \label{eq13}
E_{\rm c} & = & E_{\rm c}^{(1)} + 2\int_{R_{\rm c}}^{\infty}\left[R - a(R)\right]^2V(R)
\nonumber \\
& & {} \times \left\{\left[R - a(R)\right] W(R)-X(R)\right\} {\rm d}R,
\end{eqnarray}

\noindent where $E_{\rm c}^{(1)}$ is the first order correction

\begin{equation} \label{eq14}
  E_{\rm c}^{(1)} = -(R_{\rm c} - a_{\rm c})^2W_{\rm c} + 2(R_{\rm c} - a_{\rm c})X_{\rm c}-2Y_{\rm c}
\end{equation}

\noindent where the subscripts denote that the quantities are evaluated at $R=R_{\rm c}$ and

\begin{equation} \label{eq15}
  W(R) = \int^R V(R) {\rm d}R,
\end{equation}

\begin{equation} \label{eq16}
  X(R) = \int^R W(R) {\rm d}R,
\end{equation}

\noindent and

\begin{equation} \label{eq17}
  Y(R) = \int^R X(R) {\rm d}R.
\end{equation}

\noindent When examining the corrections at $R=R_{\rm c}$ we assume that the potential is attractive for $R>R_{\rm c}$. The correction $E_{\rm c}$ is clearly negative so that $a_{\rm c}$ approaches $a$ from above as illustrated in figure 2. If the potential is a negative series of inverse powers then $V_{\rm c}<0$, $W_{\rm c}>0$, $X_{\rm c}<0$ and $Y_{\rm c}>0$; if also either $a(R)<0$ and $R>R_{\rm c}$ or $R>a(R)>R_{\rm c}$ then the integrand in the last part of Eq.~(\ref{eq13}) is negative and $E_{\rm c}<E_{\rm c}^{(1)}$. Hence

\begin{equation} \label{eq18}
  a = a_{\rm c} + E_{\rm c} < a_{\rm c} + E_{\rm c}^{(1)},
\end{equation}

\noindent and therefore the first order corrected scattering length approaches $a$ from above. This is illustrated in figure 2 by the curve $a^{({\rm U})}$. The first order correction $E_{\rm c}^{(1)}$ is seen from Eq.~(\ref{eq14}) to be negative. Hence the first order corrected scattering length, $a_{\rm c} + E_{\rm c}^{(1)}$, is a better approximation than is $a_{\rm c}$. Both $a_{\rm c}$ and $a_{\rm c} + E_{\rm c}^{(1)}$ are upper bounds.

\subsection{Higher Order Corrections}

We rewrite Eq.~(\ref{eq13}) as

\begin{equation} \label{eq19}
  E_{\rm c} = E_{\rm c}^{(1)} + {\mathcal E}_{\rm c},
\end{equation}

\noindent where

\begin{widetext}
\begin{equation} \label{eq20}
  {\mathcal E}_{\rm c} = 2\int_{R_{\rm c}}^{\infty} \left[R - a(R)\right]^2V(R)\left\{\left[R - a(R)\right]W(R)-X(R)\right\} {\rm d}R
\end{equation}
\end{widetext}

\noindent represents higher order corrections. It is desirable to obtain a lower bound to the scattering length. We do this by partially taking account of the higher order corrections to obtain an expression that is correct to at least first order but also includes sufficient higher order terms to be a lower bound. We integrate Eq.~(\ref{eq20}) by parts to find

\begin{widetext}
\begin{equation} \label{eq21}
  {\mathcal E}_{\rm c} = {\mathcal E}_{\rm c}^{(2)} + \int_{R_{\rm c}}^{\infty} \left[R - a(R)\right]V(R)\left\{\left[R - a(R)\right]X(R)-2Y(R)\right\} {\rm d}R,
\end{equation}
\end{widetext}

\noindent where ${\mathcal E}_{\rm c}^{(2)}$ takes account of some, but not all, of the second order terms and is given by

\begin{widetext}
\begin{equation} \label{eq22}
  {\mathcal E}_{\rm c}^{(2)} = -(R_{\rm c}-a_{\rm c})^3W_{\rm c}^2 + 3(R_{\rm c}-a_{\rm c})^2W_{\rm c}X_{\rm c} - 2(R_{\rm c}-a_{\rm c})(X_{\rm c}^2+W_{\rm c}Y_{\rm c}) + 2X_{\rm c}Y_{\rm c}.
\end{equation}
\end{widetext}

\noindent The integrand in Eq.~(\ref{eq21}) is positive and hence ${\mathcal E}_{\rm c}>{\mathcal E}_{\rm c}^{(2)}$. Therefore

\begin{equation} \label{eq23}
  a> a_{\rm c} + E_{\rm c}^{(1)} + {\mathcal E}_{\rm c}^{(2)}
\end{equation}

\noindent and the right hand side of inequality Eq.~(\ref{eq23}) is a lower bound. The expression $E_{\rm c}^{(1)} + {\mathcal E}_{\rm c}^{(2)}$ can be rewritten from Eqs.~(\ref{eq14}) and (\ref{eq22}), as

\begin{equation} \label{eq24}
  E_{\rm c}^{(1)} + {\mathcal E}_{\rm c}^{(2)} = \frac{\displaystyle -2Y_{\rm c}+2(R_{\rm c}-a_{\rm c})X_{\rm c}-(R_{\rm c}-a_{\rm c})^2W_{\rm c}}{\displaystyle 1+X_{\rm c}-(R_{\rm c}-a_{\rm c})W_{\rm c}}~+~\ldots,
\end{equation}

\noindent where $+~\ldots$ denotes terms of order three and more in the potential strength. The terms $E_{\rm c}^{(1)} + {\mathcal E}_{\rm c}^{(2)}$ constitute a guaranteed lower bound to the correction $E_{\rm c}$. The first term (the ratio) on the right hand side of Eq.~(\ref{eq24}) is not a guaranteed lower bound but, on the assumption that $R_{\rm c}$ is sufficiently large that the higher order terms are negligible it is, in practice, a lower bound. We have achieved bounds on the scattering length; from Eq.~(\ref{eq14}) we have an upper bound

\begin{equation} \label{eq25}
  a^{({\rm U})} = a_{\rm c} + E_{\rm c}^{(1)},
\end{equation}

\noindent and from Eq.~(\ref{eq24}) we have a lower bound

\begin{equation} \label{eq26}
  a^{({\rm L})} = a_{\rm c} + E_{\rm c}^{(1)} + {\mathcal E}_{\rm c}^{(2)},
\end{equation}

\noindent where $E_{\rm c}^{(1)} + {\mathcal E}_{\rm c}^{(2)}$ is taken as the leading term (the ratio) on the right hand side of Eq.~(\ref{eq24}).

\vspace{1.0cm}

\begin{figure}[h]
\centerline{\scalebox{0.35}{\includegraphics{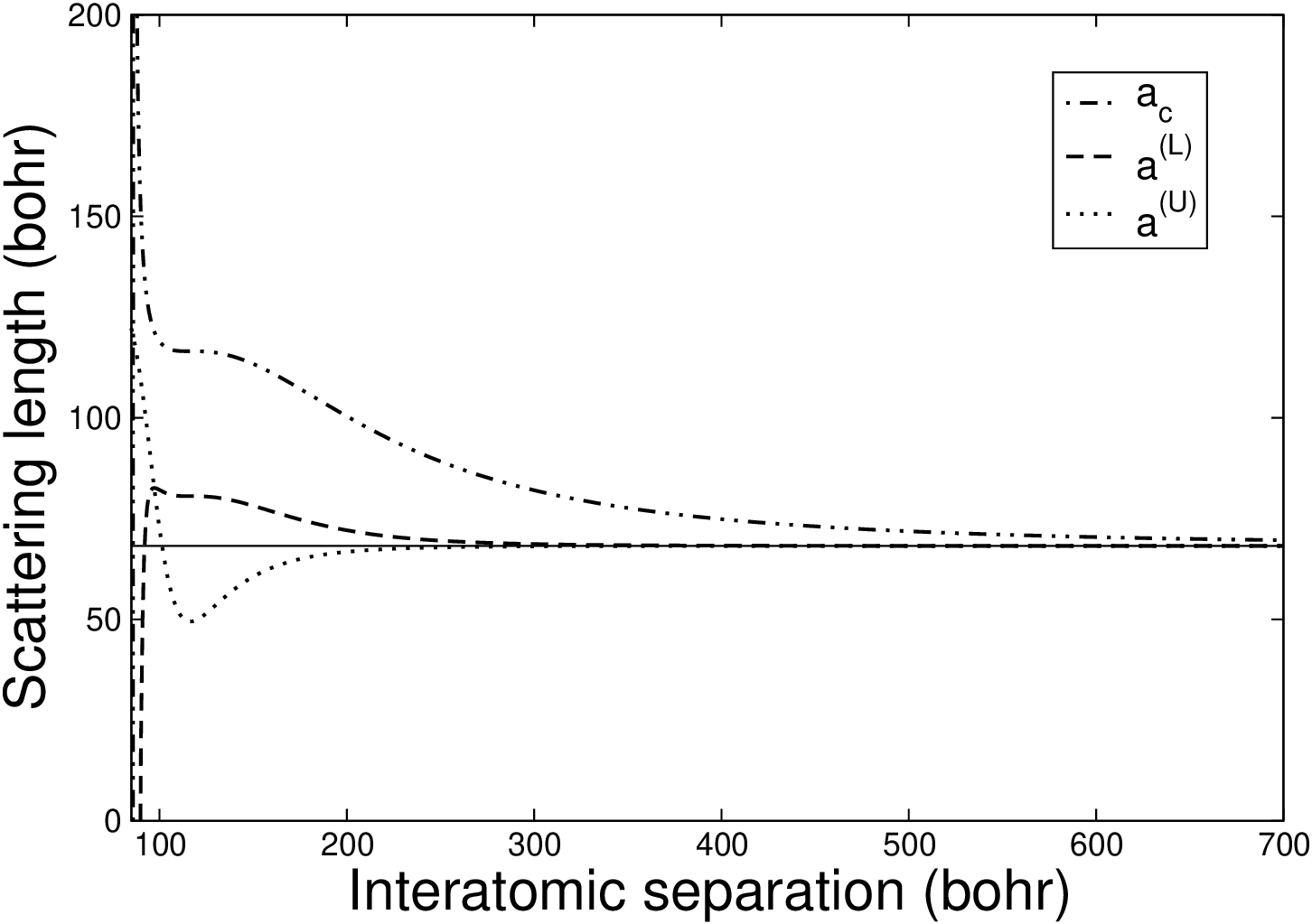}}}
\caption{The accumulated scattering length, $a_{\rm c}$, and corrected scattering lengths, $a^{({\rm L})}$ and $a^{({\rm U})}$, as functions of interatomic separation $R$.}
\end{figure}

Marinescu \cite{MAR94} derived a series of corrections to the scattering length for differing orders of the potential strength. He derived corrections from an iterative solution to a sequence of differential equations, one correction for each order of the potential strength. On substituting the functions defined by Eqs.~(\ref{eq15}), (\ref{eq16}) and (\ref{eq17}) into Marinescu's first order correction we find that it is the same as the correction given by Eq.~(\ref{eq26}). Marinescu showed that his higher order corrections alternate in sign giving upper and lower bounds. Our equations (\ref{eq25}) and (\ref{eq26}) provide bounds very simply; while they are guaranteed correct only to the first order they provide good, useful corrections for calculations with a moderately small value of $R_{\rm c}$. In the model caesium collision problem discussed in section 3.3, range of 1250 bohr proved sufficient to yield seven figure accuracy. This range agrees with that found by Marinescu. The higher order corrections are complicated. The curve $a^{({\rm L})}$ in figure 2 illustrates how the lower bound corrections leads to a scattering length that approaches $a$ from below.

\subsection{Inverse Power Series Potentials}

For the calculations shown in figures 1 and 2, we used the asymptotic potential

\begin{equation} \label{eq27}
  {\mathcal V}(R) = -C_6R^{-6} - C_8R^{-8} - C_{10}R^{-10},
\end{equation}

\noindent with the coefficients shown in section 3.3. With such inverse power potentials Eqs.~(\ref{eq14}), (\ref{eq24}) and (\ref{eq26}) assume simple forms.

When the term is $R^{-6}$ is considered alone we find that Eq.~(\ref{eq14}) yields the expression that Szmytkowski \cite{SZM95} derived from analysis of the asymptotic wavefunctions. In his analysis Szmytkowski gave also an expression for the scattering length in the form of a fraction involving Bessel functions. By replacing the Bessel functions by their asymptotic expansions for small arguments we find that Szmytkowski's ratio agrees with the lower bound $a^{({\rm L})}$ in Eq.~(\ref{eq26}).

When a term $-C_nR^{-n}$ dominates the potential the upper and lower bounds are

\begin{equation} \label{eq28}
  \tilde {a}^{({\rm U})} = \tilde {a}_{\rm c} - \tilde{\alpha}_n^{n-2} \left(\frac{\displaystyle 1}{\displaystyle n-3}~-~\frac{\displaystyle 2\tilde {a}_{\rm c}}{\displaystyle n-2}~+~\frac{\displaystyle \tilde {a}_{\rm c}^2}{\displaystyle n-1}\right),
\end{equation}

and

\begin{equation} \label{eq29}
  \tilde {a}^{({\rm L})} = \tilde {a}_{\rm c} - ~\frac{\displaystyle \tilde{\alpha}_n^{n-2} \left(\frac{\displaystyle 1}{\displaystyle n-3}~-~\frac{\displaystyle 2\tilde {a}_{\rm c}}{\displaystyle n-2}~+~\frac{\displaystyle \tilde {a}_{\rm c}^2}{\displaystyle n-1}\right)}{\displaystyle 1-\tilde{\alpha}_n^{n-2} \left(\frac{\displaystyle 1}{\displaystyle n-2} - ~\frac{\displaystyle \tilde {a}_{\rm c}}{\displaystyle n-1}\right)},
\end{equation}

\noindent where the tilde notation indicates that lengths are replaced by the dimensioneless quantities obtained by dividing by $R_{\rm c}$, and $\alpha_n$ is a length characteristic of the potential given by

\begin{equation} \label{eq30}
  \alpha_n = \left(\frac{\displaystyle 2\mu C_n}{\displaystyle \hbar^2}\right)^{1/(n-2)}.
\end{equation}

\noindent Eq.~(\ref{eq28}) agrees with our previous study \cite{JAM03} and, for $n=6$, with the analysis of Hinckelmann and Spruch \cite{HIN71}. The correction to be made to $a^{({\rm U})}$, given by Eq.~(\ref{eq20}), is asymptotically,

\begin{equation} \label{eq31}
  \tilde {{\mathcal E}}^{(U)} = \tilde {{\mathcal E}}_{\rm c} = -2~\frac{\displaystyle \tilde {\alpha}_n^{2n-4}}{\displaystyle (n-2)(2n-5)},
\end{equation}

\noindent and the correction for $a^{({\rm L})}$ is

\begin{equation} \label{eq32}
  \tilde {{\mathcal E}}^{({\rm L})} = \frac{\displaystyle \tilde {\alpha}_n^{2n-4}}{\displaystyle (n-2)(n-3)(2n-5)} = -~\frac{\displaystyle \tilde {{\mathcal E}}^{({\rm U})}}{\displaystyle 2(n-3)}.
\end{equation}

\noindent From Eq.~(\ref{eq32}) we can construct a better approximation. In these asymptotic conditions it is

\begin{equation} \label{eq33}
  a' = \frac{\displaystyle a^{({\rm U})} + 2(n-3)a^{({\rm L})}}{\displaystyle 2n-5}.
\end{equation}

\noindent We illustrate this in figure 3 for n=6.

\vspace{1.0cm}

\begin{figure}[h]
\centerline{\scalebox{0.35}{\includegraphics{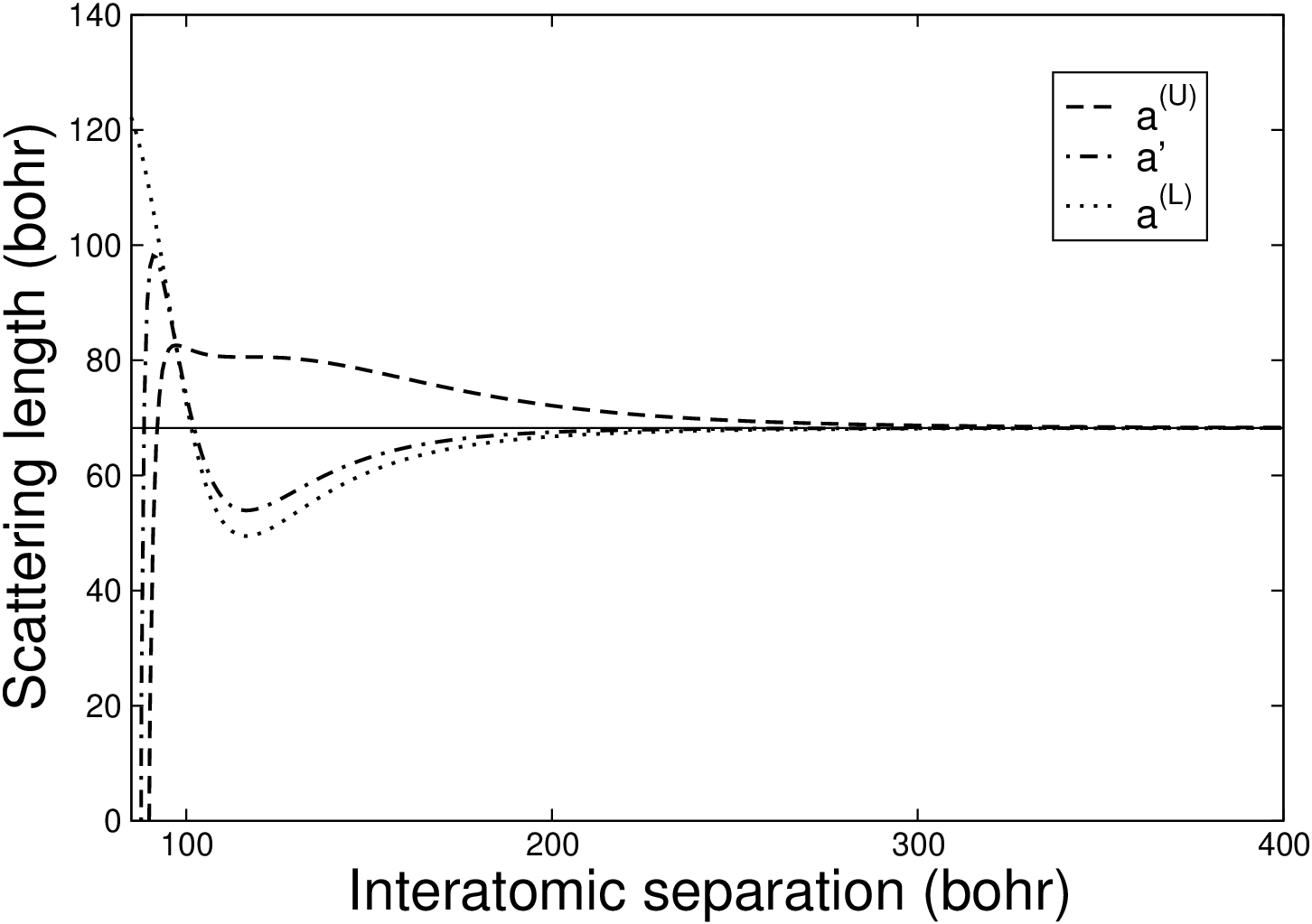}}}
\caption{The corrected scattering lengths $a'$, $a^{({\rm L})}$ and $a^{({\rm U})}$, as functions of interatomic separation $R$.}
\end{figure}

\section{Secular Perturbation Theory}

Secular perturbation theory has been applied to motion in the long range polarisation potential $-C_4R^{-4}$ by Cavagnero \cite{CAV94}. He showed how the scattering length is altered by long range interactions in accord with the predictions of Spruch {\it et al} \cite{SPR60}. We indicate below how secular perturbation theory yields the lower bound $a^{(L)}$ for the scattering length for a van der Waals $-C_6R^{-6}$ potential. The result is easily generalised for a potential such as that in Eq.~(\ref{eq27}).

We are interested in the solution of Eq.~(\ref{eq1}), with $V(R) = -C_6R^{-6}$, for small values of the wavenumber $k$ and large values of the separation $R$. Substituting $R^{-1}y(k,R)=\sqrt{(k/z)}M(z)$, where $z=kR$, in Eq.~(\ref{eq1}) we obtain

\begin{equation} \label{eq34}
  \left[z^2~\frac{\displaystyle {\rm d}^2}{\displaystyle {\rm d}z^2}~ + z~\frac{\displaystyle {\rm d}}{\displaystyle {\rm d}z}~+ z^2 -~ \frac{\displaystyle 1}{\displaystyle 4} \right] M(z) = -~\frac{\displaystyle 2\mu}{\displaystyle \hbar^2}~ \frac{\displaystyle k^4C_6}{\displaystyle z^4}~M(z).  
\end{equation}

\noindent If the right hand side of Eq.~(\ref{eq34}) were absent, $M(z)$ would be a Bessel function of order 1/2. We replace $M(z)$ by the series

\begin{equation} \label{eq35}
  M(z) = \sum_{n=-\infty}^{\infty} c_nJ_{\nu+2n}(z),
\end{equation}

\noindent and, on the right hand side of Eq.~(\ref{eq34}), we substitute \cite{ABR72}

\begin{equation} \label{eq36}
  z^{-4} J_{\nu+2n}(z) = \sum_{m=n-2}^{n+2} V_{m,n}(\nu) J_{\nu+2m}(z),
\end{equation}

\noindent where $V_{m,n}(\nu)$ are expansion coefficients, to find linear algebraic equations for $c_n$

\begin{equation} \label{eq37}
  \sum_{m=n-2}^{n+2}\!\left\{\left[\left(\nu+2m\right)^2 -~\frac{\displaystyle 1}{\displaystyle 4}\right]\delta_{m,n} + ~ \frac{\displaystyle 2\mu}{\displaystyle \hbar^2}~V_{m,n}(\nu)\right\} c_n = 0,
\end{equation}

\noindent where we have truncated the sum in expansion (\ref{eq35}) at $n=\pm n_{max}$. Eq.~(\ref{eq37}) leads to a determinantal equation satisfied by $\nu$. We include a term in $k^4$ in $\nu$ and write \cite{SAD00}

\begin{equation} \label{eq38}
  \nu = \frac{\displaystyle 1}{\displaystyle 2}~+ bk^4,
\end{equation}

\noindent and, with this given value of $\nu$, we calculate the coefficients $c_n$ from Eq.~(\ref{eq37}) after arbitrarily setting $c_0$ to unity. The equation with $n=0$ provides a new value for $\nu$

\begin{equation} \label{eq39}
  \nu^2 = \frac{\displaystyle 1}{\displaystyle 4}~-\frac{\displaystyle 2\mu}{\displaystyle \hbar^2}~V_{0,0}(\nu)~-\frac{\displaystyle 2\mu}{\displaystyle \hbar^2}~\sum_{n=-n_{max}}^{n_{max}} V_{0,n}(\nu)~(1-\delta_{n,0}) c_n.
\end{equation}

\noindent Thus we set up an iterative scheme to be solved for $\nu$ and the coefficents $c_n$. With an expansion of order 50, the low energy phase shifts for the model caesium scattering problem were determined with a matching radius of only 50 bohr \cite{CAV94b}.

The secular perturbation expansion for the scattering length appropriate for the potential in Eq.~(\ref{eq27}) is identical to that of Marinescu \cite{MAR94} to the first two orders in the secular expansion. When the leading term dominates, the secular expansion yields the correction of Szmytkowski \cite{SZM95}.

\section{Conclusion}

The variable phase method for potential scattering yields a simple first order differential equation satisfied by a function that tends in the limit of infinite interatomic separation to the scattering length \cite{CAL67}. The equation can be solved by specially adapted numerical methods, thus providing a method to compute the scattering length. This, and all other computations, being finite necessarily approximate the scattering length by its value computed at some maximum separation $R_{\rm c}$. The differential equation is readily employed to derive corrections arising from long range interactions over the range $[R_{\rm c},\infty]$. We derived two such corrections accurate to at least first order which provide upper and lower computed bounds to the scattering length. In a model collision problem these corrections provide a 40-fold reduction in the value of $R_{\rm c}$ that is necessary to let us obtain an accurate scattering length. The corrections are very simple and we suggest that use of the first order correction with a moderate value of $R_{\rm c}$ is sufficient. Other methods involve a more complete but more complicated correction, such as the $50^{th}$ order secular perturbation theory \cite{CAV94,CAV94b} or higher order expansions of the solution of the zero-energy Schr\"odinger equation \cite{MAR94}, applied to a calculation with a much smaller value of $R_{\rm c}$. However, with techniques such as interval doubling \cite{JAM03,COT95} there is little advantage in having $R_{\rm c}$ very small. We note that the first order differential equation for the effective range \cite{CAL67}, while not useful numerically, does yield the first order long range correction derived previously \cite{JAM03,HIN71}.

\section*{Acknowledgments}

This work was supported by the Engineering and Physical Sciences Research Council and by the Institute for Theoretical Atomic and Molecular Physics (ITAMP). ITAMP is supported by a grant from the National Science Foundation to Harvard University and the Smithsonian Institution.


\begin{thebibliography}{}
\bibitem{KET96} W. Ketterle and N. G. van Druten, Adv. At. Mol. Opt. Phys. {\bf 37} 181 (1996)
\bibitem{JUL93} P. S. Julienne, A. M. Smith and K. Burnett, Adv. At. Mol. Opt. Phys. {\bf 30} 141 (1993)
\bibitem{WEI99} J. Weiner {\it et al}, Rev. Mod. Phys. {\bf 71} 1 (1999)
\bibitem{MOT65} N. F. Mott and H. S. W. Massey, {\it The Theory of Atomic Collisions} (Oxford: Clarendon, 1965)
\bibitem{GUT84} G. Guti\'errez, M. de Llano and W. C Stwalley, Phys. Rev. B {\bf 29} 5211 (1984)
\bibitem{CAL67} F. Calogero, {\it Variable Phase Method Approach to Potential Scattering} (New York: Academic Press, 1967)
\bibitem{ORL99} T. Orlikowski, G. Staszewska and L. Wolniewicz, Mol. Phys. {\bf 96} 1445 (1999)
\bibitem{JAM03} M. J. Jamieson {\it et al}, J. Phys. B: At. Mol. Opt. Phys. {\bf 36} 1085 (2003)
\bibitem{CAV94} M. G. Cavagnero, Phys. Rev. A {\bf 50} 2841 (1994)
\bibitem{HIN71} O. Hinckelmann and L. Spruch, Phys. Rev. A {\bf 3} 642 (1971)
\bibitem{MAR94} M. Marinescu, Phys. Rev. A {\bf 50} 3177 (1994)
\bibitem{SZM95} R. Szmytkowski, J. Phys. A: Math. Gen. {\bf 28} 7333 (1995)
\bibitem{CHE61} C. W. Clenshaw {\it et al}, {\it Modern Computing Methods} (London: Her Majesty Stationary Office, 1961)
\bibitem{JOH73} B. R. Johnson, J. Comput. Phys. {\bf 13} 445 (1973)
\bibitem{JOH77} B. R. Johnson, J. Chem. Phys. {\bf 67} 4086 (1977)
\bibitem{MAN93} D. E. Manolopoulos {\it et al}, J. Comput. Phys. {\bf 105} 169 (1993)
\bibitem{FRI94} R. S. Friedman and M. J. Jamieson, Comput. Phys. Commun. {\bf 85} 231 (1994)
\bibitem{MAN95} D. E. Manolopoulos and S. K. Gray, J. Chem. Phys. {\bf 102} 9214 (1995)
\bibitem{GRI93} G. F. Gribakin and V. V. Flambaum, Phys. Rev. A {\bf 48} 546 (1993)
\bibitem{BLA67} J. M. Blatt, J. Comput. Phys. {\bf 1} 382 (1967)
\bibitem{COT95} R. C\^ot\'e and M. J. Jamieson, J. Comput. Phys. {\bf 118} 388 (1995)
\bibitem{SPR60} L. Spruch, T. F. O'Malley and L. Rosenberg, Phys. Rev. {\bf 5} 375 (1960)
\bibitem{ABR72} M. Abramowitz and I. A. Stegun {\it Handbook of Mathematical Functions} (New York: Dover, 1972)
\bibitem{SAD00} H. R. Sadeghpour {\it et al}, J. Phys. B: At. Mol. Opt. Phys. {\bf 33} R93 (2000)
\bibitem{CAV94b} M. J. Cavagnero, unpublished
\end{thebibliography}
\end{document}